\documentclass[pra,superscriptaddress]{revtex4}
\usepackage{amssymb}
\usepackage{amsmath}

\begin{document}
\def\be{\begin{eqnarray}}
\def\ee{\end{eqnarray}}
\def\Tr{\mathrm{Tr}}

\title{Mutually unbiased measurement based entanglement witnesses}
\author{Tao Li}
\thanks{litao@btbu.edu.cn}
\affiliation{School of Science, Beijing Technology and Business University, Beijing 100048, China}
\author{Le-Min Lai}
\thanks{lailemin@outlook.com}
\affiliation{School of Mathematical Sciences,  Capital Normal University,  Beijing 100048,  China}
\author{Shao-Ming Fei}
\thanks{feishm@mail.cnu.edu.cn}
\affiliation{School of Mathematical Sciences,  Capital Normal University,  Beijing 100048,  China}
\affiliation{Max Planck Institute for Mathematics in the Sciences, Leipzig 04103, Germany}
\author{Zhi-Xi Wang}
\thanks{wangzhx@cnu.edu.cn}
\affiliation{School of Mathematical Sciences,  Capital Normal University,  Beijing 100048,  China}

\begin{abstract}
We study entanglement witness and present a construction of entanglement witnesses in terms of the mutually unbiased measurements (MUMs).
These witnesses include the entanglement witnesses constructed from mutually unbiased bases (MUBs) as a special case.
Comparing with the dimension dependence of MUBs, the witnesses can be always constructed from a complete set of $d+1$ MUMs for any dimension $d$.
We show that our witness can detect entanglement better than previous separability criterion given also by MUMs. And the approach can be experimentally implemented.
\end{abstract}

\maketitle

\section{Introduction}
Quantum entanglement plays fundamental roles in in quantum information processing.
It has been shown that the entangled states are useful resources in quantum cryptography protocols, and can
be used to enhance the efficiency of quantum computing \cite{3,4}.
One of the key problems in the theory of quantum entanglement is the detection of entanglement \cite{1,2}.
There have been numerous criteria to distinguish quantum entangled states from
the separable ones, such as positive partial transposition criterion \cite{5,6,7}, realignment
criterion \cite{8,9,10,11,12,13,14}, covariance matrix criterion \cite{15}, and correlation matrix criterion \cite{16,17}.

One of the most useful approaches to characterize quantum entanglement is the
entanglement witness (EW) \cite{18}. The advantage of EWs is that EWs provide an
economic way to detect entanglement, which does not need the full information usually obtained by the full state tomography.
What one needs is only the information about the mean value of some observable for a given quantum state.
Remarkably, it turns out that any entangled state can be detected by certain EWs. Hence the
knowledge of witnesses enables us to fully classify states of composite
quantum systems.

An entanglement witness $W$ is an hermitian operator which is block-positive
but not positive. A bipartite state $\rho$ is separable if and only if $\Tr (\rho W)\geqslant0$ for all entanglement witnesses.
In Ref. \cite{19}, the authors connected the entanglement witness to mutually unbiased bases (MUBs) \cite{20} and
a class of entanglement witnesses are constructed in terms of MUBs.
Such construction reproduces many well-known examples such as the celebrated reduction map and the Choi map
together with its generalizations.

For given dimension $d$, if there is a complete set of MUBs, i.e., there exist $d+1$ MUBs, the entanglement witness can be simply constructed.
However, currently it is known that the $d+1$ MUBs can be only obtained when $d$ is a prime power.
When $d$ is the dimension of an arbitrary composite system, the maximum number of MUBs is still unknown,
which constrains the construction of EW in Ref. \cite{19}.
Recently, Kalev and Gour generalized the concept of MUBs to mutually unbiased measurements (MUMs) \cite{21}.
These measurements, containing the complete set of MUBs as a special case, need not be rank one projectors.
Unlike the dimension dependence of MUBs, there always exists a complete set of $d+1$ MUMs that can be explicitly constructed.

In this paper, we study the entanglement witness by using mutually unbiased measurements. We provide a class of entanglement witnesses
constructed in terms of mutually unbiased measurements.

\section{MUBS AND MUMS}
Let us first review some basic definitions of mutually unbiased bases and mutually unbiased measurements.
Two orthonormal bases $\mathcal{B}_1=\{|i\rangle\}^{d}_{i=1}$ and $\mathcal{B}_2=\{|j\rangle\}^{d}_{j=1}$ of $\mathbb{C}^d$ are said to be mutually unbiased if
\be
|\langle i|j\rangle|=\frac{1}{\sqrt{d}},~~~~~~\mathrm{for~~all}~~ i,j=1,2,\cdots,d.
\ee
A set of orthonormal bases $\{\mathcal{B}_1,\mathcal{B}_2,\cdots,\mathcal{B}_m\}$ in $\mathbb{C}^d$ is called a set of mutually unbiased bases
if every pair of bases in the set is mutually unbiased.
The construction of MUBs and the maximum number of MUBs with a composite number dimension is an open problem.
Even for $d = 6$, one does not know whether or not there exist four MUBs \cite{24,25,26,27}.
In \cite{21}, the concept of MUBs is generalized to MUMs.
Let $P^{(\alpha)}=\left\{P_1^{(\alpha)},\cdots,P_{d}^{(\alpha)}\right\}$ and $P^{(\beta)}=\left\{P_1^{(\beta)},\cdots,P_{d}^{(\beta)}\right\}$ be two positive
operator-valued measures (POVMs) such that
\be
&&\Tr(P_i^{(\alpha)})=\Tr(P_i^{(\beta)})=1,\\[2mm]
&&\Tr(P_i^{(\alpha)}P_j^{(\beta)})=\frac{1}{d},
\ee
and the Hilbert-Schmidt product of two elements from the same MUM is given by a parameter $\kappa$,
\be
\Tr(P_i^{(\alpha)}P_j^{(\alpha)})=\delta_{ij}\kappa+(1-\delta_{ij})\frac{1-\kappa}{d-1},
\ee
where $1/d\leqslant \kappa \leqslant 1$.

A general construction of $d+1$ MUMs has been presented
in \cite{21}. Let $\{F_{n,b}: n =1,2,\cdots d-1, b=1 ,2,\cdots d+1\}$ be a set of $d^2-1$ Hermitian, and traceless operators acting on $\mathbb{C}^d$, satisfying $\Tr(F_{n,b}F_{n^{\prime},b^{\prime}})=\delta_{nn^{\prime}}\delta_{bb^{\prime}}$. Define $d(d+1)$ operators
\be
F_{n}^{(b)}=\left\{
              \begin{array}{ll}
                F^{(b)}-(d+\sqrt{d})F_{n,b}, & n=1,2,\cdots d-1; \\[2mm]
                (1+\sqrt{d})F^{(b)}, & n=d,
              \end{array}
            \right.
\ee
where $F^{(b)}=\sum\limits^{d-1}\limits_{n=1}F_{n,b}$, $b=1,2,\cdots d+1$. Then the $d + 1$ MUMs are given by
\be \label{MUMs}
P^{(b)}_{n}=\frac{1}{d}I+tF_{n}^{(b)},
\ee
with $b=1,2,\cdots d+1$, $n =1,2,\cdots d-1$, and $t$ is so chosen such that $P^{(b)}_{n}\geqslant0$.
Any $d + 1$ MUMs can be expressed in such form. That is to say, there always exist $d +1$ MUMs for arbitrary $d$.

\section{Entanglement witnesses based on MUMs}

Generalizing the results in \cite{19} by using MUMs, we first construct a class of trace preserving positive maps.
Let $\left\{P_1^{(\alpha)},\cdots,P_{d}^{(\alpha)}\right\}$, with $\alpha=1,2,\cdots,L$, denote $L$ MUMs.
Let $\mathcal{O}^{(\alpha)}$ be a set of orthogonal rotation in $\mathbb{R}^{d}$ around the axis $\mathbf{n}=(1,1, \cdots,1)/\sqrt{d}$.

\textit{\textbf{Theorem 1.}} The following map $\Phi$ is positive and trace preserving,
\begin{eqnarray}\label{map}
\Phi X = \frac{1}{d}\mathbb{I}_d\mathrm{Tr} X-\frac{1}{d\kappa-1}\sum\limits_{\alpha=1}^{L} \sum\limits_{k,l=1}^{d}\mathcal{O}_{kl}^{(\alpha)}\mathrm{Tr}(\widetilde{X}P_{l}^{(\alpha)})P_{k}^{(\alpha)},\nonumber
\end{eqnarray}
where $\widetilde{X}=X- \frac{1}{d}\mathbb{I}_d\mathrm{Tr} (X)$.

\textit{Proof.} To prove positivity of $\Phi$ we show that for any rank-1 projector $P = |\phi\rangle\langle \phi|$, one has $\mathrm{Tr}(\Phi P)^2\leqslant \displaystyle \frac{1}{d-1}$ \cite{21+1}.
\begin{eqnarray*}
&\mathrm{Tr}&(\Phi P)^2 \\
&=&\mathrm{Tr}\left\{\frac{1}{d}\mathbb{I}-2\frac{1}{d}\frac{1}{d\kappa-1}\sum\limits_{\alpha=1}^{L} \sum\limits_{k,l=1}^{d}\mathcal{O}_{kl}^{(\alpha)}\mathrm{Tr}(\widetilde{P}P_{l}^{(\alpha)})P_{k}^{(\alpha)}+\frac{1}{(d\kappa-1)^2}\sum\limits_{\alpha,\beta=1}^{L} \sum\limits_{k,l,m,n=1}^{d}\mathcal{O}_{kl}^{(\alpha)}\mathrm{Tr}(\widetilde{P}P_{l}^{(\alpha)})P_{k}^{(\alpha)}\mathcal{O}_{mn}^{(\beta)}\mathrm{Tr}(\widetilde{P}P_{n}^{(\beta)})P_{m}^{(\beta)}
\right\}\\
&=&\frac{1}{d}-2\frac{1}{d}\frac{1}{d\kappa-1}\sum\limits_{\alpha=1}^{L} \sum\limits_{k,l=1}^{d}\mathcal{O}_{kl}^{(\alpha)}\mathrm{Tr}(\widetilde{P}P_{l}^{(\alpha)})+\frac{1}{(d\kappa-1)^2}\sum\limits_{\alpha=1}^{L} \sum\limits_{k,l,m,n=1}^{d}\mathcal{O}_{kl}^{(\alpha)}\mathcal{O}_{mn}^{(\alpha)}\mathrm{Tr}(\widetilde{P}P_{l}^{(\alpha)})\mathrm{Tr}(\widetilde{P}P_{n}^{(\alpha)})\Tr(P_{k}^{(\alpha)}P_{m}^{(\alpha)})\\
&~&+\frac{1}{(d\kappa-1)^2}\sum\limits_{\alpha\neq \beta=1}^{L}\sum\limits_{k,l,m,n=1}^{d}\mathcal{O}_{kl}^{(\alpha)}\mathcal{O}_{mn}^{(\beta)}\mathrm{Tr}(\widetilde{P}P_{l}^{(\alpha)})\mathrm{Tr}(\widetilde{P}P_{n}^{(\beta)})\Tr(P_{k}^{(\alpha)}P_{m}^{(\beta)})\allowdisplaybreaks[4]\\
&=&\frac{1}{d}+\frac{1}{(d\kappa-1)^2}\sum\limits_{\alpha=1}^{L}
\left[ \sum\limits_{k=1}^{d}\sum\limits_{l,n=1}^{d}\mathcal{O}_{kl}^{(\alpha)}\mathcal{O}_{kn}^{(\alpha)}\mathrm{Tr}(\widetilde{P}P_{l}^{(\alpha)})\mathrm{Tr}(\widetilde{P}P_{n}^{(\alpha)})\Tr(P_{k}^{(\alpha)}P_{k}^{(\alpha)})
\right.\\&&\phantom{=\;\;}
\left.
~~~~~~~~~~~~~~~~~~~~~+\sum\limits_{k\neq m=1}^{d}\sum\limits_{l,n=1}^{d}\mathcal{O}_{kl}^{(\alpha)}\mathcal{O}_{mn}^{(\alpha)}\mathrm{Tr}(\widetilde{P}P_{l}^{(\alpha)})\mathrm{Tr}(\widetilde{P}P_{n}^{(\alpha)})\Tr(P_{k}^{(\alpha)}P_{m}^{(\alpha)})
\right] \\
&=&\frac{1}{d}+\frac{1}{(d\kappa-1)^2}\sum\limits_{\alpha=1}^{L}\left[ \sum\limits_{l,n=1}^{d}\delta_{ln}\mathrm{Tr}(\widetilde{P}P_{l}^{(\alpha)})\mathrm{Tr}(\widetilde{P}P_{n}^{(\alpha)})\kappa+\sum\limits_{k\neq m=1}^{d}\sum\limits_{l,n=1}^{d}\mathcal{O}_{kl}^{(\alpha)}\mathcal{O}_{mn}^{(\alpha)}\mathrm{Tr}(\widetilde{P}P_{l}^{(\alpha)})\mathrm{Tr}(\widetilde{P}P_{n}^{(\alpha)})\frac{1-\kappa}{(d-1)}
\right]\\
&=&\frac{1}{d}+\frac{\kappa}{(d-1)^2}\sum\limits_{\alpha=1}^{L}\sum\limits_{l=1}^{d}[\Tr(\widetilde{P}P_{l}^{(\alpha)})]^2.
\end{eqnarray*}
Taking into account that
\begin{eqnarray*}
[\Tr(\widetilde{P}P_{l}^{(\alpha)})]^2=[\Tr(PP_{l}^{(\alpha)})]^2-2\frac{1}{d}\Tr(PP_{l}^{(\alpha)})+\frac{1}{d^2}
\end{eqnarray*}
and using the inequality
$$
\sum\limits_{\alpha=1}^{L}\sum\limits_{l=1}^{d}[\Tr(PP_{l}^{(\alpha)})]^2\leqslant \frac{L-1}{d}+\frac{1-\kappa+\kappa(d-1)}{d-1},
$$
we finally arrive at
\be
&&\mathrm{Tr}(\Phi P)^2\nonumber\\
&&\leqslant\frac{1}{d}+\frac{\kappa}{(d-1)^2}\left(\frac{L-1}{d}+\frac{1-\kappa+\kappa(d-1)}{d-1}+\frac{L}{d}-\frac{2L}{d}\right) \nonumber\\
&&\leqslant\frac{1}{d\kappa-1}, \nonumber
\ee
which ends the proof of positivity. The proof of trace preservation is straightforward.  $\blacksquare$

With proper orthogonal rotation, the corresponding entanglement witness is given by
\be\label{EW}
W_{\Phi}=(d\kappa-1)\sum\limits_{i,j=1}^{d}|i\rangle\langle j|\otimes \Phi |i\rangle\langle j|,
\ee
which can be written as
\be\label{EW1}
W_{\Phi}=\frac{d\kappa+L-1}{d}\mathbb{I}_d\otimes\mathbb{I}_d-\sum\limits_{\alpha=1}^{L}\sum\limits_{k,l=1}^{d}\mathcal{O}_{kl}^{(\alpha)}\overline{P}_{l}^{(\alpha)}\otimes P_{k}^{(\alpha)},
\ee
where $\overline{P}_{l}^{(\alpha)}$ denote the conjugation of $P_{l}^{(\alpha)}$.
In particular, as we can always find $L=d+1$ MUMs for any dimension $d$, we have from (\ref{EW1})
\be \label{W}
W_{\Phi}=(1+\kappa)\mathbb{I}_d\otimes\mathbb{I}_d-\sum\limits_{\alpha=1}^{d+1}\sum\limits_{k,l=1}^{d}\mathcal{O}_{kl}^{(\alpha)}\overline{P}_{l}^{(\alpha)}\otimes P_{k}^{(\alpha)}.
\ee

\textit{\textbf{Remark.}} The parameter $\kappa$ depends on the inner product of two measurement operators of an MUM, which
characterizes how close the measurement operators are to rank one projectors, i.e., to measurement operators in MUB.
When $\kappa=1$, our witness reduces to the previous one presented in \cite{20}.
The powerfulness of Eq. (\ref{W}) is due to the fact that there always exists a complete set of mutually unbiased measurements,
which is not the case for mutually unbiased bases (see the appendix).

Consider the most simple case, $\mathcal{O}^{(\alpha)}=\mathbb{I}$ and the maximally entangled pure state
$|\phi^{+}\rangle=\displaystyle\frac{1}{\sqrt{d}}\sum\limits_{i=1}\limits^{d}|ii\rangle$. We have
\be
&&\Tr(W_{\Phi}|\phi^{+}\rangle\langle\phi^{+}|)\nonumber \\
&&=\Tr\left[(\frac{d\kappa+L-1}{d}\mathbb{I}_d\otimes\mathbb{I}_d-\sum\limits_{\alpha=1}^{L}\sum\limits_{k=1}^{d}\overline{P}_{k}^{(\alpha)}\otimes P_{k}^{(\alpha)})(|\phi^{+}\rangle \langle \phi^{+}|)\right] \nonumber \\
&&= \frac{d\kappa+L-1}{d}-\sum\limits_{\alpha=1}^{L}\sum\limits_{k=1}^{d}\Tr(\overline{P}_{k}^{(\alpha)}\otimes P_{k}^{(\alpha)}|\phi^{+}\rangle \langle \phi^{+}|))\nonumber \\
&&\leqslant \frac{d\kappa+L-1}{d}-L\kappa\\
&&=\frac{(L-1)(1-d\kappa)}{d} <0,\nonumber
\ee
where the first inequality is obtained from the indices of coincidence's inequality in \cite{28}.
Thus the witness operator detects the entanglement of the state $|\phi^{+}\rangle$.

As another example, we consider the isotropic states which are locally
unitarily equivalent to a maximally entangled state mixed with white noise:
\begin{eqnarray}
\rho_{\mathrm{iso}}=\alpha|\phi^{+}\rangle \langle \phi^{+}|+\frac{1-\alpha}{d^2}\mathbb{I},
\end{eqnarray}
where $|\phi^{+}\rangle=\displaystyle\frac{1}{\sqrt{d}}\sum\limits_{i=1}\limits^{d}|ii\rangle$, $0 \leqslant \alpha <1$.
By using the $L=d+1$ MUMs and simply taking $\mathcal{O}^{(\alpha)}=\mathbb{I}$, we get
\be\label{13}
&&\Tr(W_{\Phi}\rho_{iso})\nonumber\\
&&= \Tr\left[(\frac{d\kappa+L-1}{d}\mathbb{I}_d\otimes\mathbb{I}_d-\sum\limits_{\alpha=1}^{L}\sum\limits_{k=1}^{d}\overline{P}_{k}^{(\alpha)}\otimes P_{k}^{(\alpha)})\rho_{iso}\right]\nonumber\\
&&=(1+\kappa)-(d+1)\left(\alpha\kappa+\frac{1-\alpha}{d}\right).
\ee
From (\ref{13}) we see that if $\alpha>{1}/({d+1})$, then $\Tr(W_{\Phi}\rho_{iso})<0$, and $\rho_{iso}$ must be entangled.
Therefore, $W_{\Phi}$ detects all the entanglement in isotropic states. This result coincides with the fact that $\rho_{iso}$ is
entangled for $\alpha>{1}/({d+1})$, and separable for $\alpha\leqslant{1}/({d+1})$ \cite{22}.

In Ref. \cite{23}, a separability criterion has been presented by also using the $d+1$ MUMs.
The next example shows that our witness approach can work better in entanglement detection.
We consider the case $d=3$. One has four MUMs with $\kappa=0.358$ constructed according to Eq. (\ref{MUMs}):
\be
\mathfrak{M}_1&=&\left\{
\left(
\begin{array}{ccc}
 0.333 & 0.107 i & -0.029 i \\
 -0.107 i & 0.333 & 0 \\
 0.029 i & 0 & 0.333 \\
\end{array}
\right),
\left(
\begin{array}{ccc}
0.333 & -0.029 i & 0.107 i \\
 0.029 i &0.333 & 0 \\
  -0.107 i & 0 &0.333 \\
\end{array}
\right),
\left(
\begin{array}{ccc}
0.333 & -0.078 i & -0.078 i \\
 0.078 i &0.333 & 0 \\
 0.078 i & 0 &0.333 \\
\end{array}
\right)
\right\}, \nonumber
\ee
\be
\mathfrak{M}_2&=&\left\{
\left(
\begin{array}{ccc}
 0.333 & -0.107 & 0 \\
 -0.107 & 0.333 &  -0.029 i \\
 0 & 0.029 i & 0.333 \\
\end{array}
\right),
\left(
\begin{array}{ccc}
0.333 & 0.029 & 0 \\
 0.029 &0.333 & 0.107 i \\
 0 & -0.107 i &0.333 \\
\end{array}
\right),
\left(
\begin{array}{ccc}
0.333 & 0.078 & 0 \\
 0.078 & 0.333 & -0.078 i \\
 0 & 0.078 i & 0.333 \\
\end{array}
\right)
\right\}, \nonumber
\ee
\be
\mathfrak{M}_3&=&\left\{
\left(
\begin{array}{ccc}
0.333 & 0 & -0.107 \\
 0 &0.333 & 0.0287 \\
 -0.107 & 0.0287 &0.333 \\
\end{array}
\right),
\left(
\begin{array}{ccc}
0.333 & 0 & 0.0287 \\
 0 &0.333 & -0.107 \\
 0.0287 & -0.107 &0.333 \\
\end{array}
\right),
\left(
\begin{array}{ccc}
0.333 & 0 & 0.078217 \\
 0 &0.333 & 0.078217 \\
 0.078217 & 0.078217 &0.333 \\
\end{array}
\right)
\right\}, \nonumber
\ee
\be
\mathfrak{M}_4&=&\left\{
\left(
\begin{array}{ccc}
 0.243 & 0 & 0 \\
 0 & 0.457 & 0 \\
 0 & 0 & 0.3 \\
\end{array}
\right),
\left(
\begin{array}{ccc}
 0.3 & 0 & 0 \\
 0 & 0.243 & 0 \\
 0 & 0 & 0.457 \\
\end{array}
\right),
\left(
\begin{array}{ccc}
 0.457 & 0 & 0 \\
 0 & 0.3 & 0 \\
 0 & 0 & 0.243 \\
\end{array}
\right)
\right\}. \nonumber
\ee
A general proper rotation in $\mathbb{R}^3$ preserving the direction $\mathbf{n}=(n_1,n_2,n_3)$, with $|\mathbf{n}|=1$, is given by the Rodrigues formula
\be
R(\mathbf{n},\theta)=\left(
\begin{array}{ccc}
\mathrm{cos} \theta +n_1^2(1-\mathrm{cos} \theta)&n_1n_2(1-\mathrm{cos} \theta)-n_3\mathrm{sin} \theta&n_1n_3(1-\mathrm{cos} \theta)+n_2\mathrm{sin} \theta\\
n_1n_2(1-\mathrm{cos} \theta)+n_3\mathrm{sin} \theta&\mathrm{cos} \theta +n_2^2(1-\mathrm{cos} \theta)&n_2n_3(1-\mathrm{cos} \theta)-n_1\mathrm{sin} \theta\\
n_3n_1(1-\mathrm{cos} \theta)-n_2\mathrm{sin} \theta&n_3n_2(1-\mathrm{cos} \theta)+n_1\mathrm{sin} \theta&\mathrm{cos} \theta +n_3^2(1-\mathrm{cos} \theta)
\end{array}
\right).
\ee
Hence, taking  $\mathbf{n}=\mathbf{n_*}=(1,1,1)/\sqrt{3}$, we have
\be
\mathcal{O}(\theta):=R(\mathbf{n_*},\theta)=\left(
\begin{array}{ccc}
c_1(\theta)&c_2(\theta)&c_3(\theta)\\
c_3(\theta)&c_1(\theta)&c_2(\theta)\\
c_2(\theta)&c_3(\theta)&c_1(\theta)
\end{array}
\right),
\ee
where
\be\label{15}
c_1(\theta)=\frac{2}{3}\mathrm{cos} \theta+\frac{1}{3},\nonumber\\
c_2(\theta)=\frac{2}{3}\mathrm{cos}(\theta-\frac{2\pi}{3})+\frac{1}{3},\\
c_2(\theta)=\frac{2}{3}\mathrm{cos}(\theta+\frac{2\pi}{3})+\frac{1}{3}\nonumber.
\ee
Let us consider the following $3\otimes 3$ state $\rho$,
\be
\left(
\tiny
\begin{array}{ccccccccc}
 0.29 & -0.007-0.051 i & 0.326+0.006 i & -0.049+0.101 i & 0.041+0.137 i & 0.059+0.138 i & 0.141-0.065 i & 0.012+0.052 i & 0.122-0.007i \\
 -0.007+0.051 i & 0.009 & -0.009+0.057 i & -0.016-0.011 i & -0.025+0.004 i & -0.026+0.007 i & 0.008+0.0262 i & -0.009+0.001i & -0.002+0.022 i \\
 0.326-0.006 i & -0.009-0.057 i & 0.365 & -0.052+0.114 i & 0.049+0.153 i & 0.068+0.153 i & 0.157-0.076 i & 0.014+0.059 i & 0.137-0.011i \\
 -0.049-0.101 i & -0.016+0.011 i & -0.052-0.114 i & 0.043 & 0.041-0.037 i & 0.038-0.043 i & -0.046-0.038 i & 0.016-0.013 i & -0.023-0.041i \\
 0.041-0.137 i & -0.025-0.004 i & 0.049-0.153 i & 0.041+0.037 i & 0.071 & 0.073-0.008 i & -0.011-0.076 i & 0.027+0.002 i & 0.014-0.059i \\
 0.059-0.138 i & -0.026-0.007 i & 0.068-0.153 i & 0.038+0.043 i & 0.073+0.008 i & 0.077 & -0.002-0.079 i & 0.027+0.005 i & 0.021-0.059i \\
 0.141+0.065 i & 0.008-0.026 i & 0.157+0.076 i & -0.046+0.038 i & -0.011+0.076 i & -0.002+0.079 i & 0.082 & -0.006+0.028 i & 0.061+0.024i \\
 0.012-0.052 i & -0.009-0.001 i & 0.014-0.058 i & 0.016+0.013 i & 0.027-0.002 i & 0.027-0.005 i & -0.006-0.028 i & 0.009 & 0.004-0.022i \\
 0.122+0.007 i & -0.002-0.022 i & 0.137+0.011 i & -0.023+0.041 i & 0.014+0.059 i & 0.021+0.059 i & 0.061-0.024 i & 0.004+0.0224 i
& 0.052 \\
\end{array}
\right).\nonumber
\ee
It is direct to verify that $J(\rho)-1-\kappa=-0.0085<0$, where $J(\rho)$ is defined in \cite{23}. According to
the results in \cite{23}, the entanglement of this state is not detected since $J(\rho)<1+\kappa$.

With respect to the four MUMs, $\mathfrak{M}_i$, $i=1,2,3,4$, we take four rotations (\ref{15})
given by four angles $\{\theta_1,\theta_2,\theta_3,\theta_4\}$, respectively.
Taking $\theta_1=\theta_2=\frac{\pi }{3}$ and $\theta_3=\theta _4=0$, we have the
entanglement witness,
\be
\footnotesize
W_{\phi }=\left(
\begin{array}{ccccccccc}
 0.001 & 0 & 0 & 0 & -0.018 & -0.016 & 0 & 0.016 & -0.028 \\
 0 & 0.038 & 0 & 0 & 0 &0.016 i & -0.016 & 0 & 0 \\
 0 & 0 & 0.038 & 0.016 & -0.016 i & 0 & -0.009 & 0 & 0 \\
 0 & 0 & 0.016 & 0.038 & 0 & 0 & 0 & 0.016 i & 0 \\
 -0.018 & 0 & 0.016 i & 0 & 0.001 & 0 & 0.016 i & 0 & -0.028 \\
 -0.016 & -0.016 i & 0 & 0 & 0 & 0.038 & 0 & -0.009 & 0 \\
 0 & -0.016 & -0.009 & 0 & -0.016 i & 0 & 0.038 & 0 & 0 \\
 0.016 & 0 & 0 & -0.016 i & 0 & -0.009 & 0 & 0.038 & 0 \\
 -0.028 & 0 & 0 & 0 & -0.028 & 0 & 0 & 0 & 0.001 \\
\end{array}
\right).\nonumber
\ee
It is direct to show that
$
\Tr(\rho W_{\phi })=-0.0017,
$
which verifies that the state $\rho$ is entangled.

To show that our separability criterion based on MUM is better than the one based on MUB,
we consider the case of $d=6$ in Appendix. In this case one only knows that there exist three MUBs.
It is clearly shown that the MUM based criterion can detect the entanglement of the state, while
the MUB based criterion fails.

\section{Conclusion}
We have presented a class of entanglement witnesses based on mutually unbiased measurements.
These witnesses include the ones constructed from mutually unbiased bases as a special case
that the efficiency parameter $\kappa$ of mutually unbiased measurements is equal to $1$.
Different from the dimension dependence of MUBs, there always exists a complete set of $d+1$ MUMs for any dimension $d$.
Moreover, our witness can detect entanglement better than the separability criterion given also by MUMs \cite{23}.
Our approach can be experimentally implemented.

\bigskip
{\sf Acknowledgments}~
This work is supported by the NSF of China under Grant No. 11675113, the Research Foundation for Youth Scholars of Beijing Technology and
Business University QNJJ2017-03, Scientific Research General Program of Beijing Municipal Commission of Education (Grant No.KM201810011009),
and NSF of Beijing under No. KZ201810028042.

\clearpage
\begin{widetext}
\begin{center}
\Large
  Appendix \\
  ~
\end{center}

In order to show the advantage of the MUM based criterion (\ref{W}), we consider
the following $6\otimes6$ bipartite state,
\be
\rho=\left(
       \begin{array}{cccccc}
         A & B & B & B & B & \alpha \\
         B & A & B & B & B & \alpha \\
         B & B & A & B & B & \alpha \\
         B & B & B & A & B & \alpha \\
         B & B & B & B & A & \alpha \\
         \alpha^t & \alpha^t & \alpha^t & \alpha^t & \alpha^t & a \\
       \end{array}
     \right),
\ee
where
\be
\small
A=\left(
\begin{array}{ccccccc}
 0.058 & 0 & 0 & 0 & 0 & 0 & 0 \\
 0 & 0.022 & 0 & 0 & 0 & 0 & 0 \\
 0 & 0 & 0.022 & 0 & 0 & 0 & 0 \\
 0 & 0 & 0 & 0.022 & 0 & 0 & 0 \\
 0 & 0 & 0 & 0 & 0.022 & 0 & 0 \\
 0 & 0 & 0 & 0 & 0 & 0.022 & 0 \\
 0 & 0 & 0 & 0 & 0 & 0 & 0.022 \\
\end{array}
\right),\
B=\left(
\begin{array}{ccccccc}
 0.036 & 0 & 0 & 0 & 0 & 0 & 0 \\
 0 & 0 & 0 & 0 & 0 & 0 & 0 \\
 0 & 0 & 0 & 0 & 0 & 0 & 0 \\
 0 & 0 & 0 & 0 & 0 & 0 & 0 \\
 0 & 0 & 0 & 0 & 0 & 0 & 0 \\
 0 & 0 & 0 & 0 & 0 & 0 & 0 \\
 0 & 0 & 0 & 0 & 0 & 0 & 0 \\
\end{array}
\right), \
\alpha=\left(
\begin{array}{c}
 0.036\\
 0 \\
 0 \\
 0 \\
 0 \\
 0 \\
 0 \\
\end{array}
\right),\
\nonumber
\ee
$a= 0.058$ and $\alpha^t$ is the transpose of $\alpha$.

For $d=6$, up to now one has only three MUBs \cite{24,25,26,27}:
\begin{eqnarray*}
\left(
\begin{array}{cccccc}
 1 & 0 & 0 & 0 & 0 & 0 \\
 0 & 1 & 0 & 0 & 0 & 0 \\
 0 & 0 & 1 & 0 & 0 & 0 \\
 0 & 0 & 0 & 1 & 0 & 0 \\
 0 & 0 & 0 & 0 & 1 & 0 \\
 0 & 0 & 0 & 0 & 0 & 1 \\
\end{array}
\right),
\end{eqnarray*}
\begin{eqnarray*}
\left(
\begin{array}{cccccc}
 0.408 & 0.408 & 0.408 & 0.408 & 0.408 & 0.408 \\
 0.408 & 0.408 w & 0.408 w^{2} & 0.408 & 0.408 w & 0.408 w^{2} \\
 0.408 & 0.408 w^{2} & 0.408 w & 0.408 & 0.408 w^{2} & 0.408 w \\
 0.408 & 0.408 & 0.408 & -0.408 & -0.408 & -0.408 \\
 0.408 & 0.408 w & 0.408 w^{2} & -0.408 & -0.408 w & -0.408 w^{2} \\
 0.408 & 0.408 w^{2} & 0.408 w & -0.408 & -0.408 w^{2} & -0.408 w \\
\end{array}
\right),
\end{eqnarray*}
\begin{eqnarray*}
\left(
\begin{array}{cccccc}
 0.408 & 0.408 & 0.408 & -0.408 i & -0.408 i & -0.408 i \\
 0.408 w & 0.408 w^{2} & 0.408 & -0.408 i w & -0.408 i w^{2} & -0.408 i \\
 0.408 w & 0.408 & 0.408 w^{2} & -0.408 i w & -0.408 i & -0.408 i w^{2} \\
 0.408 i & 0.408 i & 0.408 i & -0.408 & 0.408 & 0.408 \\
 0.408 i w & 0.408 i w^{2} & 0.408 i & 0.408 w & 0.408 w^{2} & 0.408 \\
0.408 i w & 0.408 i & 0.408 i w^{2} & 0.408 w & 0.408 & 0.408 w^{2} \\
\end{array}
\right),
\end{eqnarray*}
where $w=e^{\frac{2 \pi  i}{3}}$. Taking $\mathcal{O}^{(\alpha)}=\mathbb{I}$, we obtain $\Tr(W\rho)=0.68>0$ by numerical calculation. Therefore, the entanglement of the state $\rho$ is not detected.

Now we use our MUM based criterion in terms of the $7$ MUMs constructed as follows:
\begin{eqnarray*}
\small
P^{(1)}_{1}=\left(
\begin{array}{cccccc}
 0.167 & 0.102 i &  -0.014 i & -0.014 i & -0.014 i & -0.014 i \\
 0.102 i & 0.167 & 0 & 0 & 0 & 0 \\
 0.014 i & 0 & 0.167 & 0 & 0 & 0 \\
 0.014 i & 0 & 0 & 0.167 & 0 & 0 \\
 0.014 i & 0 & 0 & 0 & 0.167 & 0 \\
 0.014 i & 0 & 0 & 0 & 0 & 0.167 \\
\end{array}
\right);
P^{(1)}_{2}=\left(
\begin{array}{cccccc}
 0.167 & -0.014 i &  0.102 i & -0.014 i & -0.014 i & -0.014 i \\
 0.014 i & 0.167 & 0 & 0 & 0 & 0 \\
 0.102 i & 0 & 0.167 & 0 & 0 & 0 \\
 0.014 i & 0 & 0 & 0.167 & 0 & 0 \\
 0.014 i & 0 & 0 & 0 & 0.167 & 0 \\
 0.014 i & 0 & 0 & 0 & 0 & 0.167 \\
\end{array}
\right);
\end{eqnarray*}
\begin{eqnarray*}
\small
P^{(1)}_{3}=\left(
\begin{array}{cccccc}
 0.167 & -0.014 i &  -0.014 i & 0.102 i & -0.014 i & -0.014 i \\
 0.014 i & 0.167 & 0 & 0 & 0 & 0 \\
 0.014 i & 0 & 0.167 & 0 & 0 & 0 \\
 0.102 i & 0 & 0 & 0.167 & 0 & 0 \\
 0.014 i & 0 & 0 & 0 & 0.167 & 0 \\
 0.014 i & 0 & 0 & 0 & 0 & 0.167 \\
\end{array}
\right);
P^{(1)}_{4}=\left(
\begin{array}{cccccc}
 0.167 & -0.014 i &  -0.014 i & -0.014i &0.102 i & -0.014 i \\
 0.014 i & 0.167 & 0 & 0 & 0 & 0 \\
 0.014 i & 0 & 0.167 & 0 & 0 & 0 \\
 0.014 i & 0 & 0 & 0.167 & 0 & 0 \\
 0.102 i & 0 & 0 & 0 & 0.167 & 0 \\
 0.014 i & 0 & 0 & 0 & 0 & 0.167 \\
\end{array}
\right);
\end{eqnarray*}
\begin{eqnarray*}
\small
P^{(1)}_{5}=\left(
\begin{array}{cccccc}
 0.167 & -0.014 i &  -0.014 i & -0.014i & -0.014 i & 0.102 i \\
 0.014 i & 0.167 & 0 & 0 & 0 & 0 \\
 0.014 i & 0 & 0.167 & 0 & 0 & 0 \\
 0.014 i & 0 & 0 & 0.167 & 0 & 0 \\
 0.014 i & 0 & 0 & 0 & 0.167 & 0 \\
 0.102 i & 0 & 0 & 0 & 0 & 0.167 \\
\end{array}
\right);
P^{(1)}_{6}=\left(
\begin{array}{cccccc}
 0.167 & -0.047 i &  -0.047 i & -0.047i & -0.047 i & -0.047 i \\
 0.047 i & 0.167 & 0 & 0 & 0 & 0 \\
 0.047 i & 0 & 0.167 & 0 & 0 & 0 \\
 0.047 i & 0 & 0 & 0.167 & 0 & 0 \\
 0.047 i & 0 & 0 & 0 & 0.167 & 0 \\
 0.047 i & 0 & 0 & 0 & 0 & 0.167 \\
\end{array}
\right).
\end{eqnarray*}

\begin{eqnarray*}
\small
P^{(2)}_{1}=\left(
\begin{array}{cccccc}
 0.167 & -0.102 & 0 & 0 & 0 & 0 \\
 -0.102 & 0.167 & -0.014 i & -0.014 i & -0.014 i & -0.014 i \\
 0 & 0.014 i & 0.167 & 0 & 0 & 0 \\
 0 & 0.014 i & 0 & 0.167 & 0 & 0 \\
 0 & 0.014 i & 0 & 0 & 0.167 & 0 \\
 0 & 0.014 i & 0 & 0 & 0 & 0.167 \\
\end{array}
\right);
P^{(2)}_{2}=\left(
\begin{array}{cccccc}
 0.167 & 0.014 & 0 & 0 & 0 & 0 \\
 0.014 & 0.167 & 0.102 i & -0.014 i & -0.014 i & -0.014 i \\
 0 & -0.102 i & 0.167 & 0 & 0 & 0 \\
 0 & 0.014 i & 0 & 0.167 & 0 & 0 \\
 0 & 0.014 i & 0 & 0 & 0.167 & 0 \\
 0 & 0.014 i & 0 & 0 & 0 & 0.167 \\
\end{array}
\right);
\end{eqnarray*}
\begin{eqnarray*}
\small
P^{(2)}_{3}=\left(
\begin{array}{cccccc}
 0.167 & 0.014 & 0 & 0 & 0 & 0 \\
 0.014 & 0.167 & -0.014 i & 0.102 i & -0.014 i & -0.014 i \\
 0 & 0.014 i & 0.167 & 0 & 0 & 0 \\
 0 & -0.102 i & 0 & 0.167 & 0 & 0 \\
 0 & 0.014 i & 0 & 0 & 0.167 & 0 \\
 0 & 0.014 i & 0 & 0 & 0 & 0.167 \\
\end{array}
\right);
P^{(2)}_{4}=\left(
\begin{array}{cccccc}
 0.167 & 0.014 & 0 & 0 & 0 & 0 \\
 0.014 & 0.167 & -0.014 i & -0.014 i & 0.102 i & -0.014 i \\
 0 & 0.014 i & 0.167 & 0 & 0 & 0 \\
 0 & 0.014 i & 0 & 0.167 & 0 & 0 \\
 0 & -0.102 i & 0 & 0 & 0.167 & 0 \\
 0 & 0.014 i & 0 & 0 & 0 & 0.167 \\
\end{array}
\right);
\end{eqnarray*}
\begin{eqnarray*}
\small
P^{(2)}_{5}=\left(
\begin{array}{cccccc}
 0.167 & 0.014 & 0 & 0 & 0 & 0 \\
 0.014 & 0.167 & -0.014 i & -0.014 i & -0.014 i & 0.102 i \\
 0 & 0.014 i & 0.167 & 0 & 0 & 0 \\
 0 & 0.014 i & 0 & 0.167 & 0 & 0 \\
 0 & 0.014 i & 0 & 0 & 0.167 & 0 \\
 0 & -0.102 i & 0 & 0 & 0 & 0.167 \\
\end{array}
\right);
P^{(2)}_{6}=\left(
\begin{array}{cccccc}
 0.167 & 0.047 & 0 & 0 & 0 & 0 \\
 0.047 & 0.167 & -0.047 i & -0.047 i & -0.047 i & -0.047 i \\
 0 & 0.047 i & 0.167 & 0 & 0 & 0 \\
 0 & 0.047 i & 0 & 0.167 & 0 & 0 \\
 0 & 0.047 i & 0 & 0 & 0.167 & 0 \\
 0 & 0.047 i & 0 & 0 & 0 & 0.167 \\
\end{array}
\right).
\end{eqnarray*}

\begin{eqnarray*}
\small
P^{(3)}_{1}=\left(
\begin{array}{cccccc}
 0.167 & 0 & -0.102 & 0 & 0 & 0 \\
 0 & 0.167 & 0.014 i & 0 & 0 & 0 \\
 -0.102 & 0.014 & 0.167 & -0.014 i & -0.014 i & -0.014 i \\
 0 & 0 & 0.014 i & 0.167 & 0 & 0 \\
 0 & 0 & 0.014 i & 0 & 0.167 & 0 \\
 0 & 0 & 0.014 i & 0 & 0 & 0.167 \\
\end{array}
\right);
P^{(3)}_{2}=\left(
\begin{array}{cccccc}
 0.167 & 0 & 0.014 & 0 & 0 & 0 \\
 0 & 0.167 & -0.102 i & 0 & 0 & 0 \\
 0.014 & -0.102 & 0.167  & -0.014 i & -0.014 i & -0.014 i \\
 0 & 0 & 0.014 i & 0.167 & 0 & 0 \\
 0 & 0 & 0.014 i & 0 & 0.167 & 0 \\
 0 & 0 & 0.014 i & 0 & 0 & 0.167 \\
\end{array}
\right);
\end{eqnarray*}
\begin{eqnarray*}
\small
P^{(3)}_{3}=\left(
\begin{array}{cccccc}
 0.167 & 0 & 0.014 & 0 & 0 & 0 \\
 0 & 0.167 & 0.014 & 0 & 0 & 0 \\
 0.014 & 0.014 & 0.167  & 0.102 i & -0.014 i & -0.014 i \\
 0 & 0 & -0.102 i & 0.167 & 0 & 0 \\
 0 & 0 & 0.014 i & 0 & 0.167 & 0 \\
 0 & 0 & 0.014 i & 0 & 0 & 0.167 \\
\end{array}
\right);
P^{(3)}_{4}=\left(
\begin{array}{cccccc}
 0.167 & 0 & 0.014 & 0 & 0 & 0 \\
 0 & 0.167 & 0.014 & 0 & 0 & 0 \\
 0.014 & 0.014 & 0.167  & -0.014 i & 0.102 i & -0.014 i \\
 0 & 0 & 0.014 i & 0.167 & 0 & 0 \\
 0 & 0 & -0.102 i & 0 & 0.167 & 0 \\
 0 & 0 & 0.014 i & 0 & 0 & 0.167 \\
\end{array}
\right);
\end{eqnarray*}
\begin{eqnarray*}
\small
P^{(3)}_{5}=\left(
\begin{array}{cccccc}
 0.167 & 0 & 0.014 & 0 & 0 & 0 \\
 0 & 0.167 & 0.014 & 0 & 0 & 0 \\
 0.014 & 0.014 & 0.167  & -0.014 i & -0.014 i & 0.102 i \\
 0 & 0 & 0.014 i & 0.167 & 0 & 0 \\
 0 & 0 & 0.014 i & 0 & 0.167 & 0 \\
 0 & 0 & -0.102 i & 0 & 0 & 0.167 \\
\end{array}
\right);
P^{(3)}_{6}=\left(
\begin{array}{cccccc}
 0.167 & 0 & 0.047 & 0 & 0 & 0 \\
 0 & 0.167 & 0.047 & 0 & 0 & 0 \\
 0.047 & 0.047 & 0.167  & -0.047 i & -0.047 i & -0.047 i \\
 0 & 0 & 0.047 i & 0.167 & 0 & 0 \\
 0 & 0 & 0.047 i & 0 & 0.167 & 0 \\
 0 & 0 & 0.047 i & 0 & 0 & 0.167 \\
\end{array}
\right).
\end{eqnarray*}

\begin{eqnarray*}
\small
P^{(4)}_{1}=\left(
\begin{array}{cccccc}
 0.167 & 0 & 0 & -0.102 & 0 & 0 \\
 0 & 0.167 & 0 & 0.014 & 0 & 0 \\
 0 & 0 & 0.167  & 0.014  & 0 & 0 \\
 -0.102 & 0.014 & 0.014  & 0.167 & -0.014 i & -0.014 i\\
 0 & 0 & 0 & 0.014 i & 0.167 & 0 \\
 0 & 0 & 0 & 0.014 i & 0 & 0.167 \\
\end{array}
\right);
P^{(4)}_{2}=\left(
\begin{array}{cccccc}
 0.167 & 0 & 0 & 0.014 & 0 & 0 \\
 0 & 0.167 & 0 & -0.102 & 0 & 0 \\
 0 & 0 & 0.167  & 0.014  & 0 & 0 \\
 0.014 & -0.102 & 0.014  & 0.167 & -0.014 i & -0.014 i\\
 0 & 0 & 0 & 0.014 i & 0.167 & 0 \\
 0 & 0 & 0 & 0.014 i & 0 & 0.167 \\
\end{array}
\right);
\end{eqnarray*}
\begin{eqnarray*}
\small
P^{(4)}_{3}=\left(
\begin{array}{cccccc}
 0.167 & 0 & 0 & 0.014 & 0 & 0 \\
 0 & 0.167 & 0 & 0.014 & 0 & 0 \\
 0 & 0 & 0.167  & -0.102 & 0 & 0 \\
 0.014 &0.014 & -0.102  & 0.167 & -0.014 i & -0.014 i\\
 0 & 0 & 0 & 0.014 i & 0.167 & 0 \\
 0 & 0 & 0 & 0.014 i & 0 & 0.167 \\
\end{array}
\right);
P^{(4)}_{4}=\left(
\begin{array}{cccccc}
 0.167 & 0 & 0 & 0.014 & 0 & 0 \\
 0 & 0.167 & 0 & 0.014 & 0 & 0 \\
 0 & 0 & 0.167  &0.014 & 0 & 0 \\
 0.014 &0.014 & 0.014  & 0.167 & 0.102 i & -0.014 i\\
 0 & 0 & 0 & -0.102 i & 0.167 & 0 \\
 0 & 0 & 0 & 0.014 i & 0 & 0.167 \\
\end{array}
\right);
\end{eqnarray*}
\begin{eqnarray*}
\small
P^{(4)}_{5}=\left(
\begin{array}{cccccc}
 0.167 & 0 & 0 & 0.014 & 0 & 0 \\
 0 & 0.167 & 0 & 0.014 & 0 & 0 \\
 0 & 0 & 0.167  &0.014 & 0 & 0 \\
 0.014 &0.014 & 0.014  & 0.167 & -0.014 i & 0.102 i\\
 0 & 0 & 0 & 0.014 i & 0.167 & 0 \\
 0 & 0 & 0 & -0.102 i & 0 & 0.167 \\
\end{array}
\right);
P^{(4)}_{6}=\left(
\begin{array}{cccccc}
 0.167 & 0 & 0 & 0.047 & 0 & 0 \\
 0 & 0.167 & 0 & 0.047 & 0 & 0 \\
 0 & 0 & 0.167  &0.047 & 0 & 0 \\
 0.047 &0.047 & 0.047  & 0.167 & -0.047 i & -0.047 i\\
 0 & 0 & 0 & 0.047 i & 0.167 & 0 \\
 0 & 0 & 0 & 0.047 i & 0 & 0.167 \\
\end{array}
\right).
\end{eqnarray*}

\begin{eqnarray*}
\small
P^{(5)}_{1}=\left(
\begin{array}{cccccc}
 0.167 & 0 & 0 & 0 & -0.102 & 0 \\
 0 & 0.167 & 0 & 0 & 0.014 & 0 \\
 0 & 0 & 0.167 & 0 & 0.014 & 0 \\
 0 & 0 & 0 & 0.167 & 0.014 & 0 \\
 -0.102 & 0.014 & 0.014 & 0.014 & 0.167 & -0.014 i \\
 0 & 0 & 0 & 0 & 0.014 i & 0.166667 \\
\end{array}
\right);
P^{(5)}_{2}=\left(
\begin{array}{cccccc}
 0.167 & 0 & 0 & 0 & 0.014 & 0 \\
 0 & 0.167 & 0 & 0 & -0.102 & 0 \\
 0 & 0 & 0.167 & 0 & 0.014 & 0 \\
 0 & 0 & 0 & 0.167 & 0.014 & 0 \\
 0.014 & -0.102 & 0.014 & 0.014 & 0.167 & -0.014 i \\
 0 & 0 & 0 & 0 & 0.014 i & 0.167 \\
\end{array}
\right);
\end{eqnarray*}
\begin{eqnarray*}
\small
P^{(5)}_{3}=\left(
\begin{array}{cccccc}
 0.167 & 0 & 0 & 0 & 0.014 & 0 \\
 0 & 0.167 & 0 & 0 & 0.014 & 0 \\
 0 & 0 & 0.167 & 0 & -0.102 & 0 \\
 0 & 0 & 0 & 0.167 & 0.014 & 0 \\
 0.014 & 0.014 & -0.102 & 0.014 & 0.167 & -0.014 i \\
 0 & 0 & 0 & 0 & 0.014 i & 0.167 \\
\end{array}
\right);
P^{(5)}_{4}=\left(
\begin{array}{cccccc}
 0.167 & 0 & 0 & 0 & 0.014 & 0 \\
 0 & 0.167 & 0 & 0 & 0.014 & 0 \\
 0 & 0 & 0.167 & 0 & 0.014 & 0 \\
 0 & 0 & 0 & 0.167 & -0.102 & 0 \\
 0.014 & 0.014 & 0.014 & -0.102 & 0.167 &-0.014 i \\
 0 & 0 & 0 & 0 &0.014 i & 0.167 \\
\end{array}
\right);
\end{eqnarray*}
\begin{eqnarray*}
\small
P^{(5)}_{5}=\left(
\begin{array}{cccccc}
 0.167 & 0 & 0 & 0 & 0.014 & 0 \\
 0 & 0.167 & 0 & 0 & 0.014 & 0 \\
 0 & 0 & 0.167 & 0 & 0.014 & 0 \\
 0 & 0 & 0 & 0.167 & 0.014 & 0 \\
 0.014 & 0.014 & 0.014 & 0.014 & 0.167 & 0.102 i \\
 0 & 0 & 0 & 0 & -0.102 i & 0.167 \\
\end{array}
\right);
P^{(5)}_{6}=\left(
\begin{array}{cccccc}
 0.167 & 0 & 0 & 0 & 0.047 & 0 \\
 0 & 0.167 & 0 & 0 & 0.047 & 0 \\
 0 & 0 & 0.167 & 0 & 0.047 & 0 \\
 0 & 0 & 0 & 0.167 & 0.047 & 0 \\
 0.047 & 0.047 & 0.047 & 0.047 & 0.167 & -0.047 i \\
 0 & 0 & 0 & 0 & 0.047 i & 0.167 \\
\end{array}
\right).
\end{eqnarray*}

\begin{eqnarray*}
\small
P^{(6)}_{1}=\left(
\begin{array}{cccccc}
 0.167 & 0 & 0 & 0 & 0 & -0.102 \\
 0 & 0.167 & 0 & 0 & 0 & 0.014 \\
 0 & 0 & 0.167 & 0 & 0 & 0.014 \\
 0 & 0 & 0 & 0.167 & 0 & 0.014 \\
 0 & 0 & 0 & 0 & 0.167 & 0.014 \\
 -0.102 & 0.014 & 0.014 & 0.014 & 0.014 & 0.167 \\
\end{array}
\right);
P^{(6)}_{2}=\left(
\begin{array}{cccccc}
 0.167 & 0 & 0 & 0 & 0 & 0.014 \\
 0 & 0.167 & 0 & 0 & 0 & -0.102 \\
 0 & 0 & 0.167 & 0 & 0 & 0.014 \\
 0 & 0 & 0 & 0.167 & 0 & 0.014 \\
 0 & 0 & 0 & 0 & 0.167 & 0.014 \\
 0.014 & -0.102 & 0.014 & 0.014 & 0.014 & 0.167 \\
\end{array}
\right);
\end{eqnarray*}
\begin{eqnarray*}
\small
P^{(6)}_{3}=\left(
\begin{array}{cccccc}
 0.167 & 0 & 0 & 0 & 0 & 0.014 \\
 0 & 0.167 & 0 & 0 & 0 & 0.014 \\
 0 & 0 & 0.167 & 0 & 0 & -0.102 \\
 0 & 0 & 0 & 0.167 & 0 & 0.014 \\
 0 & 0 & 0 & 0 & 0.167 & 0.014 \\
 0.014 & 0.014 & -0.102 & 0.014 & 0.014 & 0.167 \\
\end{array}
\right);
P^{(6)}_{4}=\left(
\begin{array}{cccccc}
 0.167 & 0 & 0 & 0 & 0 & 0.014 \\
 0 & 0.167 & 0 & 0 & 0 & 0.014 \\
 0 & 0 & 0.167 & 0 & 0 & 0.014 \\
 0 & 0 & 0 & 0.167 & 0 & -0.102 \\
 0 & 0 & 0 & 0 & 0.167 & 0.014 \\
 0.014 & 0.014 & 0.014 & -0.102 & 0.014 & 0.167 \\
\end{array}
\right);
\end{eqnarray*}
\begin{eqnarray*}
\small
P^{(6)}_{5}=\left(
\begin{array}{cccccc}
 0.167 & 0 & 0 & 0 & 0 & 0.014 \\
 0 & 0.167 & 0 & 0 & 0 & 0.014 \\
 0 & 0 & 0.167 & 0 & 0 & 0.014 \\
 0 & 0 & 0 & 0.167 & 0 & 0.014 \\
 0 & 0 & 0 & 0 & 0.167 & -0.102 \\
 0.014 & 0.014 & 0.014 & 0.014 & -0.102 & 0.167 \\
\end{array}
\right);
P^{(6)}_{6}=\left(
\begin{array}{cccccc}
 0.167 & 0 & 0 & 0 & 0 & 0.047 \\
 0 & 0.167 & 0 & 0 & 0 & 0.047 \\
 0 & 0 & 0.167 & 0 & 0 & 0.047 \\
 0 & 0 & 0 & 0.167 & 0 & 0.047 \\
 0 & 0 & 0 & 0 & 0.167 & 0.047 \\
 0.047 & 0.047 & 0.047 & 0.047 & 0.047 & 0.167 \\
\end{array}
\right).
\end{eqnarray*}

\begin{eqnarray*}
\small
P^{(7)}_{1}=\left(
\begin{array}{cccccc}
 0.086 & 0 & 0 & 0 & 0 & 0 \\
 0 & 0.289 & 0 & 0 & 0 & 0 \\
 0 & 0 & 0.164 & 0 & 0 & 0 \\
 0 & 0 & 0 & 0.156 & 0 & 0 \\
 0 & 0 & 0 & 0 & 0.153 & 0 \\
 0 & 0 & 0 & 0 & 0 & 0.149 \\
\end{array}
\right);
P^{(7)}_{2}=\left(
\begin{array}{cccccc}
 0.135 & 0 & 0 & 0 & 0 & 0 \\
 0 & 0.108 & 0 & 0 & 0 & 0 \\
 0 & 0 & 0.297 & 0 & 0 & 0 \\
 0 & 0 & 0 & 0.156 & 0 & 0 \\
 0 & 0 & 0 & 0 & 0.153 & 0 \\
 0 & 0 & 0 & 0 & 0 & 0.149 \\
\end{array}
\right);
\end{eqnarray*}
\begin{eqnarray*}
\small
P^{(7)}_{3}=\left(
\begin{array}{cccccc}
 0.155 & 0 & 0 & 0 & 0 & 0 \\
 0 & 0.127 & 0 & 0 & 0 & 0 \\
 0 & 0 & 0.117 & 0 & 0 & 0 \\
 0 & 0 & 0 & 0.299 & 0 & 0 \\
 0 & 0 & 0 & 0 & 0.153 & 0 \\
 0 & 0 & 0 & 0 & 0 & 0.149 \\
\end{array}
\right);
P^{(7)}_{4}=\left(
\begin{array}{cccccc}
 0.165 & 0 & 0 & 0 & 0 & 0 \\
 0 & 0.138 & 0 & 0 & 0 & 0 \\
 0 & 0 & 0.128 & 0 & 0 & 0 \\
 0 & 0 & 0 & 0.121 & 0 & 0 \\
 0 & 0 & 0 & 0 & 0.299 & 0 \\
 0 & 0 & 0 & 0 & 0 & 0.149 \\
\end{array}
\right);
\end{eqnarray*}
\begin{eqnarray*}
\small
P^{(7)}_{5}=\left(
\begin{array}{cccccc}
 0.172 & 0 & 0 & 0 & 0 & 0 \\
 0 & 0.145 & 0 & 0 & 0 & 0 \\
 0 & 0 & 0.135 & 0 & 0 & 0 \\
 0 & 0 & 0 & 0.128 & 0 & 0 \\
 0 & 0 & 0 & 0 & 0.123 & 0 \\
 0 & 0 & 0 & 0 & 0 & 0.298 \\
\end{array}
\right);
P^{(7)}_{6}=\left(
\begin{array}{cccccc}
 0.287 & 0 & 0 & 0 & 0 & 0 \\
 0 & 0.193 & 0 & 0 & 0 & 0 \\
 0 & 0 & 0.159 & 0 & 0 & 0 \\
 0 & 0 & 0 & 0.136 & 0 & 0 \\
 0 & 0 & 0 & 0 & 0.119 & 0 \\
 0 & 0 & 0 & 0 & 0 & 0.106 \\
\end{array}
\right).
\end{eqnarray*}
By calculation with $\mathcal{O}^{(\alpha)}=\mathbb{I}$, we get $\Tr(W\rho)=-0.0114<0$. This is, the state $\rho$ is entangled. The effectiveness of (\ref{W}) is due to the fact that there always exists a complete set of mutually unbiased measurements, which is not the case for mutually unbiased bases.
\end{widetext}

\begin{thebibliography}{00}
\bibitem{3} A. K. Ekert, Phys. Rev. Lett. \textbf{67}, 661 (1991).

\bibitem{4} C. A. Fuchs, N. Gisin, R. B. Griffiths, C-S. Niu, and A. Peres, Phys. Rev. A \textbf{56}, 1163 (1997).

\bibitem{1} M. A. Nielsen and I. L. Chuang, \textit{Quantum Computation and
Quantum Information} (Cambridge University Press, Cambridge,
2000).

\bibitem{2} R. Horodecki, P. Horodecki, M. Horodecki, and K. Horodecki,
Rev. Mod. Phys. \textbf{81}, 865 (2009).

\bibitem{5} A. Peres, Phys. Rev. Lett. \textbf{77}, 1413 (1996).

\bibitem{6} M. Horodecki, P. Horodecki, and R. Horodecki, Phys. Lett. A \textbf{223}, 1 (1996).

\bibitem{7} P. Horodecki, Phys. Lett. A \textbf{232}, 333 (1997).

\bibitem{8}O. Rudolph, Phys. Rev. A \textbf{67}, 032312 (2003).

\bibitem{9} K. Chen and L.-A. Wu, Quantum Inf. Comput. \textbf{3}, 193 (2003).

\bibitem{10} M. Horodecki, P. Horodecki, and R. Horodecki, Open Syst. Inf. Dyn. \textbf{13}, 103 (2006).

\bibitem{11} K. Chen and L. A.Wu, Phys. Lett. A \textbf{306}, 14 (2002).

\bibitem{12} K. Chen and L. A.Wu,Phys. Rev. A \textbf{69}, 022312 (2004).

\bibitem{13} P. Wocjan and M. Horodecki, Open Syst. Inf. Dyn. \textbf{12}, 331 (2005).

\bibitem{14} S. Albeverio, K. Chen, and S. M. Fei, Phys. Rev. A \textbf{68}, 062313 (2003).

\bibitem{15} O. Guhne, P. Hyllus, O. Gittsovich, and J. Eisert, Phys. Rev.
Lett. \textbf{99}, 130504 (2007).

\bibitem{16} J. de Vicente, Quantum Inf. Comput. \textbf{7}, 624 (2007).

\bibitem{17} J. de Vicente, J. Phys. A: Math. Theor. \textbf{41}, 065309 (2008).

\bibitem{18} B. M. Terhal, Phys. Lett. A \textbf{271}, 319 (2000).

\bibitem{19} D. Chru$\acute{s}$ci$\acute{n}$ski, G. Sarbicki, F. Wudarski, Phys. Rev. A \textbf{97}, 032318 (2018).

\bibitem{20} W. K. Wootters and B. D. Fields, Ann. Phys. (NY) \textbf{191}, 363
(1989).

\bibitem{21} A. Kalev and G. Gour, New J. Phys. \textbf{16}, 053038 (2014).

\bibitem{21+1} I. Bengtsson and $\dot{Z}$czkowski, Geometry of Quantum States: An
Introduction to Quantum Entanglement (Cambridge University
Press, Cambridge, 2006).

\bibitem{22} R. A. Bertlmann, K. Durstberger, B. C. Hiesmayr, and
P. Krammer, Phys. Rev. A \textbf{72}, 052331 (2005).

\bibitem{23} B. Chen, T. Ma, and S. M. Fei, Phys. Rev. A \textbf{89}, 064302 (2014).

\bibitem{24} S. Brierley and S. Weigert, Phys. Rev. A \textbf{78}, 042312 (2008).

\bibitem{25} S. Brierley and S. Weigert, Phys. Rev. A \textbf{79}, 052316 (2009).

\bibitem{26} P. Raynal, X. L¡§u, and B.-G. Englert, Phys. Rev. A \textbf{83}, 062303
(2011).

\bibitem{27} D. McNulty and S.Weigert, J. Phys. A:Math. Theor. \textbf{45}, 102001
(2012).

\bibitem{28} C. Spengler, M. Huber, S. Brierley, T. Adaktylos, and B. C.
Hiesmayr, Phys. Rev. A \textbf{86}, 022311 (2012).

\end{thebibliography}
\end{document}